\documentclass[sigconf, screen]{acmart}
\usepackage{soul}
\usepackage{graphicx} 
\usepackage{natbib}

\copyrightyear{2026}
\acmYear{2026}
\setcopyright{cc}
\setcctype{by}
\acmConference[GI '26]{Graphics Interface Conference}{}

\begin{document}

\title[Investigating Mixed Reality as a Bridge Between Paper-Based and Digital Artifacts in UI/UX Design]{Stitching the Divide: Investigating Mixed Reality as a Bridge Between Paper-Based and Digital Artifacts in UI/UX Design}

\author{Abidullah Khan}
\orcid{0000-0002-2131-9497}
\affiliation{%
  \institution{Polytechnique Montréal}
  \city{Montréal}
  \country{Canada}
}
\email{abid-ullah.khan@polymtl.ca}

\author{Jinghui Cheng}
\orcid{0000-0002-8474-5290}
\affiliation{%
  \institution{Polytechnique Montréal}
  \city{Montréal}
  \country{Canada}
}
\email{jinghui.cheng@polymtl.ca}

\begin{abstract}
UI/UX designers work with both paper-based and digital artifacts but lack tools that seamlessly integrate the two. Mixed Reality (MR) offers under-explored opportunities to combine the strengths of both design environments. To examine these opportunities, we first conducted interviews with 19 professional UI/UX designers to understand their current experiences using paper and digital artifacts. Motivated and informed by the interview insights, we organized nine conceptual-probe user study sessions in which designers engaged with a MR-probe that combined paper and digital prototyping processes and brainstormed MR's potential in UI/UX design. We found that participants valued MR for enabling continuous hybrid design workflows, reducing manual reconstruction, supporting spatially anchored workspaces, and facilitating real-time cross-medium collaboration. They also envisioned future MR tools with AI assistance, richer interactive and dynamic content, and the ability to manage diverse design artifacts within a unified environment. From these findings, we derive four design dimensions for future MR systems that could enable more fluid, creative, and collaborative design practices.
\end{abstract}



\maketitle

\section{Introduction}

In a UI/UX design process, paper-based physical artifacts and digital tools play complementary yet indispensable roles. Paper artifacts such as sketches, sticky notes, and hand-drawn wireframes allow designers to rapidly externalize ideas, think divergently, and engage in collaborative discussions in co-located settings~\cite{10.1145/581710.581723, christensen_chapter_2020, gamboa_conversations_2022}. On the other hand, artifacts created in digital tools (e.g., Figma) offer precision, scalability, and shared access, enabling the refinement of concepts, version management, and effective collaboration across distributed teams~\cite{jensen_physical_2018, jonson_design_2005, feng_understanding_2023}. However, these two worlds, as well as their respective benefits, are usually separated, making it challenging for designers to achieve a seamless and coherent design process.

Prior studies have attempted to reconcile this separation of physical and digital environments. For example,~\citet{klemmer_integrating_2008} examined wall-based systems that enabled fluid transitions between traditional sketching and digital augmentation to support collaborative design.~\citet{amores_showme_2015} introduced a remote collaboration system that used immersive gestural communication to bridge physical expression and digital interaction. More recently,~\citet{drey_vrsketchin_2020} proposed \textit{VRSketchIn}, which extended sketching practices into immersive 3D spaces through pen and tablet interaction, while ~\citet{bauerova_user_2025} highlighted how virtual reality can enhance brainstorming and prototyping in team-based settings. 

Despite these advances, existing systems often focus on either digitizing paper artifacts or creating fully immersive digital environments, without properly capturing the complementary strengths of both. Mixed Reality (MR) offers unique potentials to further address this problem. Aiming to merge real and virtual environments~\cite{speicher_what_2019}, MR tools can seamlessly integrate physical and digital design environments into a unified workflow that preserves the creativity affordance of paper-based artifacts while leveraging the precision and scalability of digital tools. Prior work on MR for design mostly examined physical prototyping~\cite{zhang_wizard_2024,kent_mixed_2021}, leaving the opportunity of MR tools for UI/UX design under-explored. In this paper, we take initial steps to investigate these issues from a designer-centric perspective.

Particularly, we explore how UI/UX designers envision MR as a medium for bridging paper-based and digital design artifacts. To achieve these goals, we first conducted semi-structured interviews with 19 professional UI/UX designers to understand how they create, use, and manage paper-based and digital artifacts, and transition between the two. Insights from the interviews further motivated and informed us to explore MR-based tools as a bridging medium in a conceptual-probe user study with nine experienced designers. We created a conceptual MR probe that merged paper-based and digital prototyping processes to elicit participants' perceptions and expectations during the study. The participants discussed their perceived values of MR tools in supporting the integration of paper-based and digital design artifacts and envisioned possible extensions. Our findings revealed that designers viewed MR as a promising medium for maintaining continuity across hybrid artifacts, enabling smoother transitions between low- and high-fidelity representations, and supporting collaborative design activities.

Based on these results, we derived four design dimensions to inform future MR tools as a bridge between physical and digital environments in UI/UX design, for (1) integrating diverse and heterogeneous design artifacts, (2) blending lo-fi and hi-fi prototyping, (3) leveraging spatial anchoring, and (4) facilitating different levels of collaboration. Together, these dimensions offer general directions and concrete guidance for developing future MR tools that can enable designers to fluidly combine the unique strengths of both physical and digital environments in their workflows.
\section{Related Work}
In today's UI/UX design process, integrating physical and digital artifacts has become increasingly important~\cite{10.1145/3689433}. Despite advances in digital tools, early-stage design work still heavily depends on traditional methods like pen and paper for their immediacy and low effort~\cite{10.1145/1358628.1358811}. However, their lack of integration with digital systems often leads to fragmented workflows~\cite{10.1145/1358628.1358811}, highlighting the need for hybrid approaches that combine the strengths of both mediums. Below, we briefly review previous work related to our study, focused on (1) traditional practice and artifact workflows in UI/UX design and (2) mixed reality techniques for integrating physical and digital artifacts within the design process.

\subsection{UI/UX Design Practice and Artifact Workflows}
In the traditional UI/UX design process, physical and digital artifacts play a central role in externalizing, exploring, refining, and communicating ideas. Existing research has recognized that freehand sketches, physical prototypes, and other tangible artifacts are essential tools for ideation, because of their immediacy, flexibility, and ability to support reflective design thinking~\cite{jonson_design_2005,10.1145/765891.765986,10.1145/1375761.1375762, 10.1145/581710.581723}. For instance,~\citet{jonson_design_2005} found that while digital tools have become prevalent, designers often rely on conceptual sketches to initiate ideas, underscoring the continuing importance of low-fidelity physical artifacts in shaping design thinking. Similarly, ~\citet{10.1145/765891.765986} demonstrated that paper-based prototyping enables rapid iteration and low-cost exploration, offering affordances that are difficult to replicate in purely digital environments.

These benefits are not limited to individual ideation but extend to collaborative and reflective design practices. ~\citet{klemmer_designers_2001} highlighted how tangible interfaces can serve as shared workspaces for collaborative design, allowing multiple designers to externalize and negotiate ideas simultaneously. ~\citet{hartmann_reflective_2006} further emphasized the generative role of physical prototypes in iterative reflection, showing that integrated cycles of making, testing, and analyzing tangible artifacts enhance designers’ ability to refine and communicate concepts. ~\citet{10.1145/1375761.1375762} extended this view by framing prototypes as both filters and manifestations: they constrain exploration by focusing attention while simultaneously materializing ideas for reflection and discussion.

However, while the benefits of both physical and digital artifacts are well established, transitioning between paper and digital environments remains a persistent challenge. Many studies note that digital tools excel at precision, interactivity, and sharing, but often disconnect designers from the fluidity and richness of low-fidelity sketching~\cite{10.1145/332040.332486, 10.1145/1294211.1294233, jonson_design_2005}. Attempts to bridge this gap through hybrid paper-digital systems have demonstrated potential but remain limited as such approaches typically focus on augmenting or archiving physical artifacts rather than supporting the continuous and iterative movement of ideas across physical and digital medium~\cite{10.1145/3290605.3300917, rajaram_paper_2022,10.1145/1709886.1709894}. Consequently, designers often face fragmented workflows, needing to reconstruct, reinterpret, or manually transfer ideas when moving from early sketches to digital prototypes. Our work responds to this challenge by investigating how mixed reality (MR) environments can integrate paper-based and digital artifacts, supporting continuous ideation, iteration, and collaborative design activities without the disruption inherent in current hybrid approaches.

\subsection{Mixed Reality Tools in the Design Process}
Mixed Reality (MR), a continuum between Augmented Reality (AR) and Virtual Reality (VR), merges real and virtual environments to allow real-time interaction between physical and digital objects~\cite{speicher_what_2019}. This capability has attracted growing interest in design, where MR is seen as a means to move beyond screen-based interactions and support ideation, prototyping, and evaluation in more immersive and collaborative ways~\cite{kent_mixed_2021}. However, translating these potentials into real design practice remains difficult~\cite{ashtari_creating_2020}. 

Some researchers investigated immersive technologies to support the sketching activity that is essential for early-stage design. For example, ~\citet{ardito_vrsketch_2021} investigated 2D sketching in VR and found that adjusting hand and pen transparency can mitigate occlusion and influence sketching speed and accuracy. \citet{geyer_ideavis_2012} proposed \textit{IdeaVis} and examined hybrid paper digital support for co-located sketching sessions. They found that designers strongly prefer maintaining paper-based sketching practices because they support rapid idea externalization and rich social interaction, while also benefiting from selective digital augmentation support to improve visibility, group awareness, and the ability to revisit and reuse design artifacts without disrupting existing workflows.

Rapid prototyping is another aspect explored in previous works. For example,~\citet{bauerova_user_2025} examined team-based brainstorming and product design prototyping in virtual and physical settings. They found that while VR environments increased engagement and supported faster task completion, they also led to higher perceived complexity and lower creative output compared to pure physical environments. MR techniques have also been investigated as a medium for co-design and feedback collection. For example, \citet{el-jarn_can_2020} found that XR environments can enhance remote collaboration by enabling real-time feedback, shared presence, and collective manipulation of digital artifacts, although such tools remain underutilized in professional design workflows. 

Despite these advancements, existing research on MR in design has focused almost exclusively on physical product prototyping with limited attention to traditional UI/UX design workflows. Its potential to support UI/UX design has not yet been fully investigated in the literature. Our work addresses this gap by investigating how MR can integrate physical and digital artifacts to support the UI/UX design process.

\section{Preliminary Study to Understand the Current Usage of Paper and Digital Artifacts}
To explore how UI/UX designers currently use and manage paper and digital design artifacts in their workflows, we conducted a semi-structured interview study with 19 professional UI/UX designers. Below, we describe the study methods, present key results, and discuss the main insights that shaped our subsequent research. This study has been approved by the research ethics board of our institution.

\subsection{Methods}
We recruited 19 professional UI/UX designers through LinkedIn posts, email outreach, and personal connections. The participant group consisted of nine women and ten men, ranging in age from their 20s to 40s. The participants had a variety of design-related roles including UI/UX designer, product designer, and graphic designer. They came from diverse educational backgrounds such as architecture, industrial design, computer science, business, and cognitive neuroscience. They had also worked on key projects across domains including financial technology, e-commerce, educational tools, enterprise systems, and mobile applications. Together, the participants reflected a diverse range of experience in their current UI/UX design profession: eight participants had 1–2 years of experience, six had 3–6 years, and five had more than 7 years of experience, including one senior professional with 15 years in the field.

We conducted semi-structured interviews to explore how designers use and organize paper and digital artifacts during their design process. The interviews were structured into three sections. The first section gathered demographic and background information. The second section focused on their day-to-day practices for creating and working with paper and digital artifacts. We asked about the tools they used, the types of artifacts they created, and how they organized, stored, or reused these artifacts. The third section explored how participants transitioned between paper and digital mediums and the challenges they faced during this process. All interviews were conducted remotely using Zoom or Microsoft Teams and lasted between 45 and 60 minutes. Each participant was compensated with \$30 CAD for their time.

Interview recordings were transcribed using Microsoft Word's automatic transcription tool and manually reviewed for accuracy. We then conducted a thematic analysis~\cite{vaismoradi_content_2013, aronson_pragmatic_1995} to identify common patterns, practices, and challenges related to the types of artifacts created, their management, their perceived advantages and disadvantages, and the challenges associated with transitioning between mediums. Transcripts were coded using Atlas.ti\footnote{https://atlasti.com}. The analysis was conducted by two researchers, who each independently coded approximately half of the dataset. We began with open coding to identify relevant segments of text, and iteratively refined these initial codes into broader categories and themes. The researchers then compared and discussed their codes and emerging themes, resolving differences through iterative discussion and reaching consensus on the final coding structure. The process continued until no new themes emerged. Finally, we synthesized the results using an affinity diagram constructed in Miro to visualize connections between themes and organize our key insights.

\subsection{Results}
Our analysis revealed the following themes about the tension between paper-based and digital artifacts, problems caused due to the lack of continuity across these mediums, and the manual strategies designers use to bridge them.

\subsubsection{Tension Between Exploratory and Structured Design Practices:} 
Participants described a fundamental tension between paper-based and digital artifacts in their design workflows. Paper-based artifacts supported rapid, unconstrained exploration, allowing designers to externalize ideas freely without being limited by specific tool structures. As P5 shared, ``\textit{Paper and pen works the most because I feel like I have the flexibility to change what I want... even if it's crazy, I can understand what is there.}'' P13 also explained, ``\textit{Every time we used pen and paper, we ended up with a better final design.}'' In contrast, many participants felt that high-fidelity digital environments emphasized visual polish too early, discouraging open-ended exploration, forcing designers to revert to paper during early ideation. As P7 explained, ``\textit{Sometimes I feel like I can't think too much outside the box... It can hold me back from exploring other alternatives. At those times, I usually try to take a step back and turn to the physical side just to break out of that constraint.}'' Rather than serving interchangeable roles, paper-based and digital artifacts created a tension between exploratory flexibility and structured refinement, requiring designers to continuously shift between two fragmented mediums.

\subsubsection{Lost Ideas and Workflows due to the Lack of Continuity Across Mediums:}
While paper artifacts offered flexibility and creative freedom during early ideation, many participants found them difficult to organize as their design process evolved. Sketches were often scattered across different pages, notebooks, or sticky notes, leading to confusion and lost ideas. As P5 described, ``\textit{I know sometimes you lose papers. So the idea is like forgotten to be put on the digital place.}'' This lack of organization often disrupted workflows and, more critically, prevented ideas from being transferred into digital tools, where they could be preserved, refined, and further developed. In contrast, many digital tools provide rich organization and version control features. For example, P11 stated, ``\textit{Figma and Adobe XD have a very nice versioning system where you can save a state of a file... So I can rollback, see an early version, or just continue with the latest one.}'' However, these benefits did not extend to paper-based artifacts, resulting in isolated and fragmented design workflows.

\subsubsection{Strategies for Maintaining Continuity Between Paper-based and Digital Artifacts}

To address the lack of integration between paper-based and digital artifacts, participants developed a range of manual strategies. These include the following.

\paragraph{Transferring Paper Artifacts into Digital Formats:}
A common strategy our participants mentioned was transferring paper artifacts into digital spaces. This not only helped preserve and organize their initial ideas but also made collaboration and further development easier. For example, P19 explained an extensive approach, stating, ``\textit{After drawing the wireframes on paper, I take a photo of each screen and arrange the pages accordingly. [...] I'll organize them in Figma. Once arranged, I share the Figma file with other developers for further collaboration.}'' Some participants shifted to digital sketching as a strategy to overcome the challenges of transitioning between paper-based and digital artifacts, avoiding the overhead of scanning and manually transferring paper sketches. For instance, P5 described ``\textit{So that's when I think even I shifted to using the iPad... because when you're trying to sketch it and then scan it, it becomes a problem.}''

\paragraph{Tagging and Labelling Paper Artifacts:}
A few participants also developed personal strategies for marking paper artifacts with contextual cues to maintain continuity when moving between physical and digital workflows. These included writing dates to track design progress over time, adding screen names or titles to identify each page, and grouping related sketches together by content or layout. For example, P4 explained, ``\textit{Every paper should have a name. For example, the homepage or the profile page. On the back of the page, we write the name to keep everything organized.}'' Such practices allowed participants to revisit paper-based sketches without losing ideas or context when switching back from digital tools.

\subsection{Implications for MR Design Tools}
\label{sec:pre_insights}
Findings from our preliminary study point to a broader need for tools that can better support continuity across physical and digital design activities. Rather than treating paper-based and digital artifacts as separate stages of the workflow, future systems could enable designers to fluidly combine, preserve, and revisit artifacts across mediums without interrupting their creative process. Mixed Reality (MR) presents a promising direction for supporting such interactions by enabling physical and digital design elements to coexist and be manipulated within a shared environment.
\section{Conceptual-Probe User Study to Explore the Potential of MR as a Bridging Medium in UI/UX Design}
To further investigate how MR technologies could support the integration of paper-based and digital artifacts while maintaining continuity across physical and digital design activities, we conducted a user study to understand designers' perceptions and expectations of such tools. In order to encourage designers to envision how this emerging technology could augment design practices and move beyond constraints imposed by existing tools and processes, we developed an MR conceptual probe that served as a concrete representation of how such integration could be realized in practice, helping participants visualize its potential in design work and inspire new ideas and features for future MR-based design tools.

\subsection{Methods}
This study has been approved by the research ethics board of our institution. Below, we describe the participants, the creation of the MR conceptual probe, the study procedure, and the data collection and analysis methods.

\begin{table*}[t]
\centering
\small
\caption{Summary of Conceptual-Probe Study Participants}
\label{tab:study-participant-info}
\resizebox{\textwidth}{!}{%
\begin{tabular}{ccllll}
\toprule
\textbf{ID} & \textbf{Gender} & \textbf{Yrs of Exp.} & \textbf{Key Projects} & \textbf{Education/Background} \\ \midrule
P1  & Female & 1 year & Medical web app, Design site & Masters in Industrial Engineering \\ 
P2  & Male & 6 years   & Visual design, Brand identity, Website redesign & PhD in Software and Computer Engineering \\ 
P3  & Female & 2 years   & Pizzeria simulator, Graphics tool, Fitness app & Bachelor in Software Engineering \\ 
P4  & Female & 3 years   & GenAI tool, Consulting apps, Banking systems & Masters in Software and Computer Engineering \\ 
P5  & Female & 1 year & Startup websites, Music player app & Bachelor in Industrial Design \\ 
P6  & Male & 12 years  & Map web app, Train systems & Masters in Industrial Engineering \\ 
P7  & Female & 2 years   & Charity education website, Eye-tracking SPL study & Masters in Software Engineering \\ 
P8  & Female & 1 year   & Dental project, Streaming prototype & Masters in UX \\ 
P9  & Female & 1 year   & NUX prototype, Marketing study & Masters in UX \\ 
\bottomrule
\end{tabular}%
}
\end{table*}

\subsubsection{Participants}
We recruited nine professional UI/UX designers to participate in our study through posts on professional platforms such as LinkedIn and through personal connections. The sample size is determined based on data saturation achieved during the analysis; nine participants were sufficient to provide rich qualitative insights for the study. To be eligible, participants were required to have at least one year of professional experience in UI/UX design and to have completed at least one UI/UX related design project. The final group included seven women and two men, aged between their 20s and 40s, with design experience ranging from one to twelve years. Participants received an incentive of \$60 CAD for completing the study. Prior exposure to immersive technologies was limited, with only one participant reporting experience with VR gaming. Table ~\ref{tab:study-participant-info} summarizes the characteristics of our participants.

\begin{figure*}[t]
    \centering
    \includegraphics[width=\textwidth]{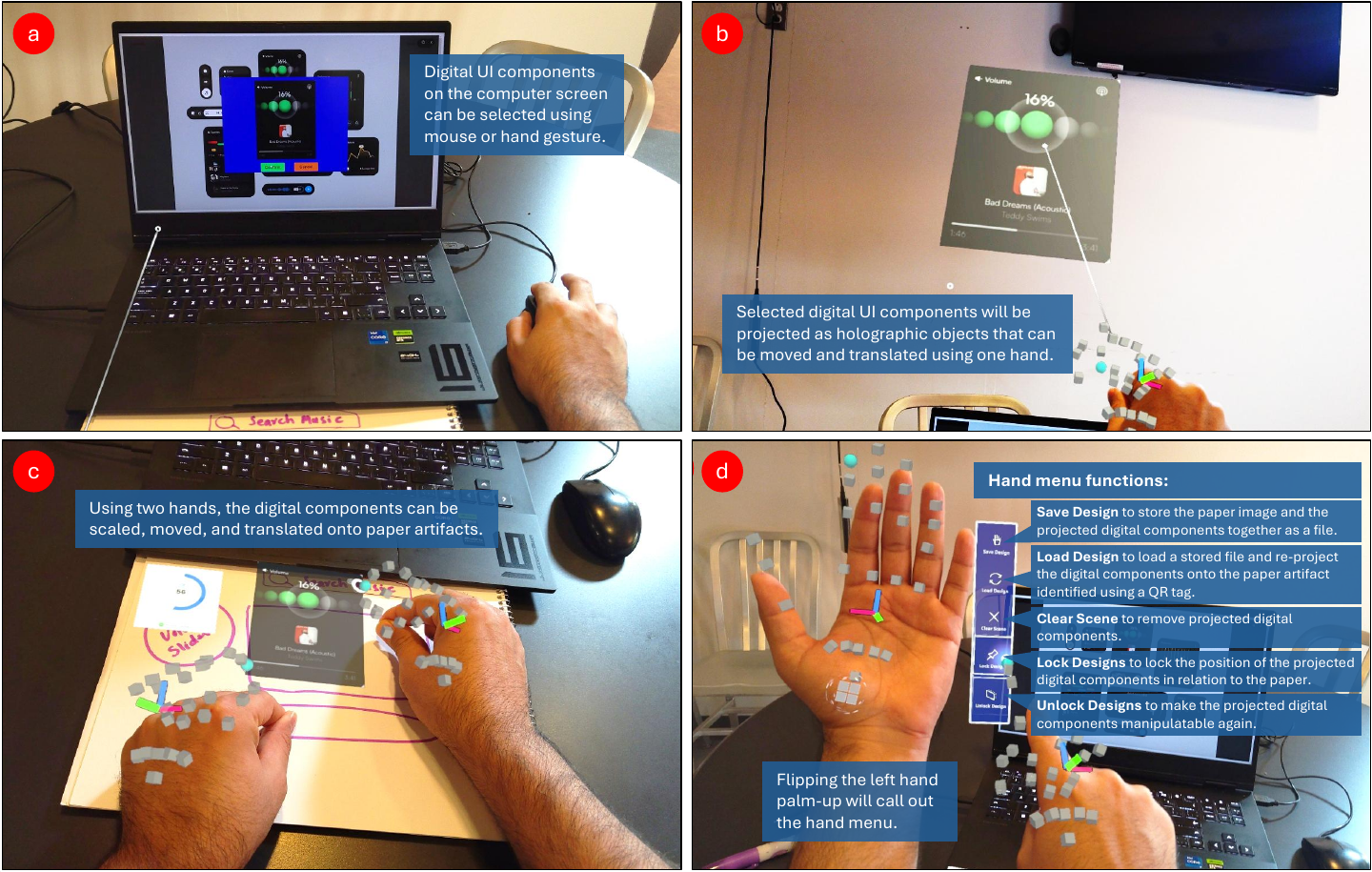}
    \Description{Photos showing the features offered by the MR conceptual probe: browsing, interacting, manipulating and placing, and locking digital components. In (a) the designer browses an online design library on a laptop screen and selects a digital component using a mouse. In (b) the selected component appears as a hologram in the physical environment through the HoloLens. In (c) the designer uses two hands to scale the hologram and place it on the paper sketch. In (d) the designer uses the hand menu to lock the digital component.}
    \caption{Illustration of tasks supported by the MR conceptual probe. The images were taken through HoloLens 2. (a) The user can view a design library on their laptop screen through the headset and select digital UI components to be brought into the physical environment using the mouse or hand gestures. (b) After confirming, the selected component appears as a holographic object in the physical workspace. (c) Using hand gestures, the user can resize the projected artifact and position it on the paper sketch. (d) The user can lock the placed digital components, save the mixed-reality artifact, and retrieve it using the hand menu.}
    \label{fig:mr_manipulation}
\end{figure*}

\subsubsection{The Creation of the Conceptual Probe for Mixed-Reality Prototyping}
The probe was developed as an MR design system to demonstrate how a mixed paper and digital prototyping workflow can be created with MR technology. The system is intentionally kept simple to facilitate reflection and idea generation, rather than prescribing solutions. With the MR system, digital artifacts can be integrated into a paper-based sketching process. The paper and the digital artifacts can then be preserved together as \textit{mixed-reality artifacts} that exist simultaneously in both physical and digital worlds. The system was developed on the HoloLens~2\footnote{https://www.microsoft.com/en-ca/hololens/} and implemented in Unity3D\footnote{https://unity.com/} using C\#, with MRTK2\footnote{https://github.com/microsoft/MixedRealityToolkit-Unity} for hand gestures. Users wearing the HoloLens~2 headset can see the physical papers, the digital design artifacts displayed on a laptop screen, as well as holographic design artifacts projected in the physical space. They could use the system to complete the following tasks:

\textbf{(1) Bringing a digital UI component into the physical environment.} When sketching ideas on paper, users may wish to draw inspiration from online design libraries or incorporate standardized UI components from an existing design system to enrich their concepts. To do that, users could use either a mouse or hand gestures to select digital components from a laptop screen. (Figure~\ref{fig:mr_manipulation}a shows a music player UI component being selected from the screen.) Once selected, the digital component will pop out from the screen and become a holographic object in the physical environment (Figure~\ref{fig:mr_manipulation}b).

\textbf{(2) Manipulating the digital component in the physical environment.} Once instantiated, users can directly manipulate the digital design artifacts within the physical environment. Using two-handed gestures, users can scale digital elements, and with one or both hands, they can translate and position them relative to existing paper sketches. (Figure~\ref{fig:mr_manipulation}c shows that the music player component is manipulated and put on the paper sketch, together with another digital component.) By allowing direct interaction with both paper and digital elements in a single environment, the system helps maintain the ideas exist across both mediums, avoiding the need to switch between them.

\textbf{(3) Preserving the mixed-reality design artifacts.} Once the digital component is placed, users can lock it from the hand menu (activated by flipping the left hand palm-up) to prevent unintended selection or movement (Figure~\ref{fig:mr_manipulation}d). This way, the digital components are merged with the paper sketch to form a mixed-reality design artifact. From the same hand menu, users can save the mixed-reality artifact into their digital design library; each of such artifact includes a picture of the paper artifact, screenshots of the digital components, and a JSON configuration file that stores the related positions of these elements. Such interaction helps ideas stay within the context without getting lost across the mediums.

\textbf{(4) Retrieving the mixed-reality design artifacts from both physical and digital environments.} Once saved, a mixed-reality artifact can be retrieved and further manipulated in both the physical and the digital environments. In the digital environment, the artifact can be reconstructed from the JSON configuration file and displayed as an image on the screen. In the physical environment, the paper sketch will be identified with a QR tag and the digital components can be re-projected on the sketch based on the JSON configuration file. From the hand menu (Figure~\ref{fig:mr_manipulation}d), users can then choose to further manipulate the existing components or remove them. They may also put more digital components on the artifact.

These features, although intentionally simple, demonstrate a vision of an MR system that allows designers to integrate digital artifacts with paper-based sketches, seamlessly transition between sketching, exploring and reusing elements from digital libraries, preserving design states, revisiting prior design alternatives, and refining their concepts without breaking their creative flow.

\subsubsection{Procedure}
The study was conducted in person, with each participant completing an approximately two-hour individual study session. Each session was structured into the following three phases: 

\textbf{(1) Exploration Phase}, where we briefly discussed the concept of using MR to combine paper and digital design artifacts and participants explored our MR probe for about 30 minutes. During this exploration, we first provided a brief tutorial on the features of the probe. Participants were then asked to imagine a scenario of designing an online learning platform and wear the HoloLens~2 headset to use the MR probe during the ideation process. They were encouraged to sketch on paper or a whiteboard and use the MR system to integrate their sketches with digital elements found online. Participants were asked to think aloud during this process and share their thoughts and opinions on how MR could support the integration between paper-based and digital artifacts, as well as the general idea of using MR in design.

\textbf{(2) Ideation Phase}, where participants engaged in a brainstorming activity for about 20 minutes to propose new ideas and features for future MR tools, particularly focusing on how such tools could support integration between paper-based and digital design artifacts, as well as broader applications in UI/UX design. At the beginning of the brainstorming session, participants were encouraged to reflect on their design workflow that involved both paper and digital design artifacts. They were provided sketchbooks, sticky notes, and a whiteboard to capture and explain their ideas. 

\textbf{(3) Discussion Phase}, where participants were asked to reflect on the ideas they had developed during the brainstorming phase and were asked about both the opportunities and challenges of integrating physical and digital design environments in the design process using MR.

\subsubsection{Data Collection and Analysis}
All study sessions were recorded, including the MR view through the HoloLens device portal to capture participants' interactions during the exploration phase. The recordings were transcribed using Microsoft Word's transcription tool and manually checked for accuracy. We then conducted a thematic analysis~\cite{vaismoradi_content_2013, aronson_pragmatic_1995} to identify themes related to the potential and values of MR for bridging paper-based and digital design artifacts in UI/UX design, as well as their desired features and functionalities for such tools. The first author led the analysis, with frequent discussions and input from the second author. We began with open coding to capture relevant parts of the transcripts and then iteratively refined these codes into broader themes. We used ATLAS.ti\footnote{https://atlasti.com} to support coding and Miro to build an affinity diagram that helped us organize the themes and synthesize the main insights.

\subsection{Results}
Participants were generally intrigued and excited by the idea of using MR to integrate paper-based and digital artifacts. They discussed the perceived values of such tools and proposed ideas for desired features. We organized the themes identified from our analysis around these two main topics.
 
\subsubsection{Potentials and Values of MR for Bridging Paper-Digital Design Workflows in UI/UX Design:}
Participants reflected on their experiences with the conceptual probe and discussed the envisioned potential and values of future MR tools that could support paper-based and digital design workflows in UI/UX design. We organized these perceived and envisioned potentials and values into the following themes.

\textit{\textbf{Facilitating Hybrid Design and Ideation Without Breaking Workflow Continuity} ($N=8$):}
Our participants emphasized the value of MR tools in the early phases of design, where quick sketches on paper can be integrated with digital content for creative exploration without requiring the need to shift into fully digital environments. This enabled designers to continue exploring ideas iteratively while selectively incorporating digital UI elements. As P2 reported, ``\textit{I think it definitely helps the ideation phase, like I can rapidly explore and test different ideas.}'' In addition, participants also highlighted the potential of MR tools to support a more playful and engaging approach to brainstorming and experimentation with evolving ideas, allowing designers to combine and manipulate ideas without disrupting their creative flow. For instance, P1 explained, ``\textit{I would have, say, a whiteboard, then go online and try to find some images I could work with... Then I’d put them on the whiteboard and go ahead and sketch or annotate on top. And if I have a series of buttons that I like, I can just plug them in and see if they work.}'' Moreover, some participants also reflected on the envisioned potential of MR tools that preserve the flexibility and iterative nature of paper-based artifacts, while allowing designers to seamlessly transition into digital platforms when needed. As P6 described, ``\textit{Being able to do creation here and then being unlimited in the way you create stuff, and then only then using a design tool such as Figma to do the specification---that would allow you to do divergent design on paper and then convergent design in the specification tool on your computer.}'' Such interactions eliminate the need to prematurely transition into fully digital tools.

\textit{\textbf{Eliminating Manual Reconstruction Between Paper-based and Digital Artifacts} ($N=6$):}
Our participants emphasized the importance of efficiency in design, particularly given the time pressures that often limit opportunities for creative exploration. They saw MR tools as a way to eliminate the need for manual effort involved in transferring, recreating, and refining paper-based ideas in digital environments, which allowed designers to directly manipulate digital content on top of their paper-based sketches, reducing the risk of losing ideas between mediums and quickly exploring multiple alternatives. For instance, P3 mentioned, ``\textit{It's faster for me to create a mockup, like a low-fidelity wireframe, on paper because I can just use my pen or something, and it's quicker. I can't do the same thing as easily in Figma. But once I already have this [MR tool], I can just grab the elements and place them.}'' Moreover, some participants envisioned MR as a way to close the gap between coming up with ideas and materializing ideas, by enabling a more direct progression from early sketches to structured design artifacts without intermediate reconstruction steps. They noted that traditional tools often slow down the translation of complex design concepts into usable artifacts. P6 highlighted this challenge, stating, ``\textit{The strength would be in relying on that hybrid mode where you design something really quickly, and then instantly it becomes a library of objects. Because one of the frustrating parts of your experience is that there is a gap between what's in your mind and what you are able to design, especially when you're designing complex components and complex behaviours.}''

\textit{\textbf{Preserving and Integrating Feedback Across Paper and Digital Artifacts} ($N=6$):}
Our findings indicated that feedback and iteration are essential in collaborative design, yet current tools often feel limited and inefficient. Feedback provided in one medium was not easily preserved or carried forward into others, often requiring designers to manually reconstruct comments or rely on memory when transitioning between paper and digital tools. Participants felt MR tools could enable a continuous flow between physical and digital representations of their work which would allow ideas to move back and forth across design environments and enable a more engaging feedback process. For example, P2 stated this, ``\textit{Being able to draw something in a physical environment and then turn it digital---but still have it in your hand [is very useful]. You're taking it to a manager, to a colleague, to anyone. They put some comments on it. You bring it back and make it physical again. You work on that physical thing again, and then make it digital again. So it's a very, very cool thing.}'' At the same time, some participants also highlighted the persistent difficulties of communicating ideas clearly through physical sketches, particularly in remote settings, where feedback is harder to capture, share, and revisit across tools. P4 discussed this, ``\textit{Most of the ideation in the first stages of design, they're working on paper, and it's so hard. And it's a lot of steps to show each other and to discuss the ideas on paper. So if we could use this kind of system to just show our colleagues our paper sketches, it would be so, so interesting and useful.}''

\textit{\textbf{Spatially Anchored Workspaces for Preserving Ideas and Context} ($N=5$):}
Our participants discussed that MR tools could provide greater flexibility in exploring and preserving ideas in spatially organized workspace, beyond what is possible with traditional sketching or digital tools. They particularly imagined the ability to use the physical workspace as an active design surface, where digital artifacts could coexist with paper sketches, avoiding the need to preserve and organize paper-based and digital artifacts manually while also reducing the risk of losing ideas during translation between environments. For example, P4 described, ``\textit{It's better than having a mood board on my laptop because I have this spatial environment... I can move around, walk, and have different ideas placed in different parts of the space... That's how I usually think---I walk! So, to have them placed around my room or office and look at them like that would be really interesting.}'' Supporting similar viewpoint, participants also discussed the potential of MR tools to create dynamic and customizable spatial MR workspaces where both paper-based and digital ideas could co-exist and grow together. P5 stated, ``\textit{So maybe mapping space to different parts of the design process---like instead of printing things and putting them on a wall or whiteboard, you could have a mood boarding wall or a brand references wall in MR... Maybe you could customize your space differently for each type of project. The same space could be used for two different purposes, and you can really customize both. I think that would be really interesting.}''

\textit{\textbf{Enabling Real-Time Cross-Medium Shared Workspaces} ($N=5$):}
Participants discussed MR as a synchronous medium that extends physical co-presence into remote settings, enabling designers to see and manipulate both paper-based and digital artifacts in real time within a shared spatial environment. As P1 explained, ``\textit{Collaborative work would be so much better---especially remotely... It's partially virtual, partially physical... You're in a meeting using the HoloLens, and everyone else can see what you're seeing... If you're sketching on paper, others could see your sketch in real time... They could annotate on your sketch or make modifications virtually.}'' Expanding on this idea, participants further discussed the potential of MR tools to create dynamic shared workspaces where designers could easily capture, share, and iterate on ideas together across both paper-based and digital workflows. P2 illustrated this, ``\textit{Maybe if we are in a design studio... Everyone in that design lab has the headset... We are in a shared environment. I have my corner... Let's imagine this is a whiteboard... While I'm working, I take a snapshot, like a card, and go to that corner and say, `OK, what do you think about this?' And they can be like, `No, man, this is not good.'}''

\subsubsection{Desired Features and Functionalities for MR-Based Design Tools}
Participants discussed and proposed a wide range of ideas and features of MR tools for supporting UI/UX design, going beyond the probe that we provided. We present the prominent ideas in this section.

\textit{\textbf{AI-Assisted Design Analysis and Collaborative Support in MR} ($N=7$):} 
Participants envisioned AI as a key component of future MR tools, particularly for reducing the effort of translating rough ideas from paper to digital artifacts. They suggested that AI could analyze hand-drawn sketches and propose relevant UI components directly within the MR environment, allowing designers to build on early ideas without manually recreating them. P8 and P9 illustrated this, ``\textit{It would be nice if it could analyze the sketch... for example, if I'm drawing a search bar, then propose something, give me suggestions... if I'm doing something like a connection page, the system could suggest some elements to add.}'' 
Some participants also expressed their desire for AI to be more like a collaborative agent acting like a teammate in the immersive design environment, where designers can interact with it alongside their paper and digital artifacts. Such interaction could reduce context switching and support seamless, integrated ideation without interrupting their workflow. P7 stated this, ``\textit{Because it's mixed reality, I think it would be a good option if I have an agent as a visual... The agent could be like your colleague, and you can collaborate with them---kind of like AI behind it but... there should be an avatar, so you can talk to them and discuss ideas.}''

\textit{\textbf{Supporting Content Organization Beyond UI Artifacts to Preserve Context} ($N=6$):}
Our participants envisioned MR as a platform that goes beyond handling static UI components, enabling designers to preserve and integrate diverse design materials such as sticky notes and diagrams within a single spatial workspace. They emphasized that such tools could reduce the risk of losing early ideas when transitioning between early ideation and later design stages. As P4 explained, ``\textit{When we work as a team, we brainstorm online... We have a lot of sticky notes. It would be great if I could have those placed right next to my sketch... Then I could sketch based on that idea without needing to rewrite them.}''

\textit{\textbf{Supporting Interactive and Dynamic UI Behaviour} ($N=5$):} 
Our participants mentioned that design work in MR should not remain limited to static sketches or images. Some envisioned MR tools that could support animations, videos, and visual effects that would show richer design ideas and communicate them more dynamically. For example, P3 explained, ``\textit{I'd like it not just to be a picture. I'd prefer it to be an animated video or an animated image, so it feels more dynamic.}''
Others saw MR’s potential to transform paper-based sketches into executable artifacts, where designers can directly encode behaviours and conditions, eliminating the need to switch to separate digital tools to manually recreate interactive experiences and behaviours. For example, P6 mentioned, ``\textit{It would be super cool if I could grab them and they become an interactive object. For example, you draw a sketch and you grab it, and it becomes an interactive sketch. And you could specify a few conditions, like the start state, end state, disabled state, active state, etc.}''

\section{Discussion}
In this study, we examined the potential of Mixed Reality (MR) tools for enabling UI/UX designers to integrate paper and digital design artifacts during their workflows. From semi-structured interviews with 19 professional UI/UX designers, we collected knowledge about how they currently create, use, organize, and transition between paper and digital artifacts. These insights motivated and informed a conceptual-probe user study, in which an MR-based probe was used to elicit designers' perceptions and expectations towards MR tools. During the study, participants reflected on the values of MR tools for design and identified opportunities for future tools. Together, our study revealed four key design dimensions that outline the considerations for future MR tools to support UI/UX workflows. We discuss these design dimensions and their practical implications in detail below.

\subsection{Dimension 1: Integrating Diverse and Heterogeneous Design Artifacts}

Participants envisioned MR tools as a medium to manage and integrate a wide spectrum of design artifacts, spanning paper sketches, sticky notes, diagrams, mood boards, and digital content such as UI screenshots, icons, animations, and videos. While static artifacts like images were valued for their intuitiveness and manipulability, participants also highlighted the potential of dynamic, interactive content to enrich early ideation, reveal interaction cues, and support exploration of complex behaviours. Importantly, designers imagined MR environments where these heterogeneous artifacts could coexist spatially with paper-based sketches, enabling a continuous evolution from low-fidelity ideas to high-fidelity wireframes or interactive components without interrupting the creative workflow.

These perspectives suggest a design opportunity for MR tools to rethink how diverse artifacts can coexist and interact within the same workspace. However, this advantage comes with potential cognitive overhead---tracking multiple physical and digital artifacts simultaneously may increase complexity if not carefully managed. Therefore, the challenge for MR tools is not only to bring heterogeneous and multi-model content together, but also to manage the different forms of attention and creativity they afford~\cite{10.1145/3654777.3676470, 10.1145/3743049.3748552}. While structured artifacts such as interactive wireframes provide clarity and consistency, they can feel rigid compared to the flexibility of early sketches and ideation notes. Multimedia elements add richness and inspiration, yet risk overwhelming the workspace if not carefully managed. Future MR tools could include context-aware visibility of rich media, allowing visually busy content (e.g., videos and animations) to hide when not needed. Also, future MR tools should have the ability to keep informal artifacts linked to later prototypes, so that diagrams, mood boards, and notes continue to inform design decisions rather than being left behind. 

\subsection{Dimension 2: Blending Lo-Fi and Hi-Fi Prototyping}

Our participants discussed different levels of prototyping fidelity and the traditional divide between low-fidelity and high-fidelity that can be integrated by MR. In conventional processes, low-fidelity sketches are confined to paper or simple digital tools, while high-fidelity interactive prototypes require significant time and effort to construct. MR could bridge this separation by allowing rough sketches to be enhanced with interactivity, dynamic behaviours, and embedded states directly in the same environment. For example, a hand-drawn button could immediately gain tappable functionality or simulate animations, allowing designers to test interactions without fully committing to a high-fidelity design.

These perspectives highlight that the central challenge for MR tools is not only to support both low- and high-fidelity prototyping, but to enable ideas to transition between them without losing their distinct strengths~\cite{10.1145/1375761.1375762}. This challenge closely aligns with the work presented by ~\citet{10.1145/1124772.1124959}, who critique the conventional assumption that fidelity exists along a single linear spectrum between low- and high-fidelity representations. Instead, they argue that prototypes can simultaneously combine different levels of visual refinement, interactivity, functionality, and data richness, resulting in mixed-fidelity artifacts. Viewed through this lens, MR tools could allow paper-based artifacts to remain visually sketch-like while still supporting interactive behaviours, dynamic feedback, and behavioural realism in real time. In this way, MR tools could make content fidelity reversible and contextual---in MR, sketches can gain interactive qualities without losing their roughness, and detailed prototypes can be simplified again to spark fresh thinking. Thus, MR could introduce ``hybrid artifacts'' that function partially as sketches and partially as interactive simulations. The complexity lies in balancing roughness with precision, creating space for open exploration while still guiding idea refinement. This suggests that fidelity in MR should be introduced in a controlled and reversible manner rather than being permanently embedded in artifacts. MR systems therefore need to support fidelity as a layered structure, where visual representations, interaction behaviours, and functional details can be independently adjusted while remaining linked to the same design artifact.

\subsection{Dimension 3: Leveraging the Benefit of Spatial Anchoring}
Participants saw MR as offering a unique advantage over traditional digital tools by anchoring design artifacts directly in the physical environment. Designers described mapping prototypes, wireframes, and reference materials to walls, desks, or defined zones, allowing them to walk around, explore, and retrieve content using spatial and environmental cues. Such anchoring was seen as supporting organization (e.g., mood boards in one area, wireframes in another) and externalizing mental models to reduce the cognitive load of tracking multiple artifacts on a 2D screen. However, such spatially anchored environments introduce potential challenges. Physical space cluttered with too many digital content could create complexity and the current space layout may limit the effectiveness of anchored content. Extended use of physical anchoring could also cause fatigue from repetitive physical interactions. MR tools therefore need to balance immersion and practicality, to ensure that spatial organization enhances rather than hinders design thinking.

To address these challenges, future MR tools could implement multi-layered spatial canvases, where design artifacts exist in conceptual layers that are only revealed when designers shift attention physically or via gestures. Moreover, such tools should support adaptive clustering, where related artifacts can be grouped automatically based on content similarity or project phase, reducing visual clutter while maintaining contextual links. MR should also consider multi-space collaboration, where multiple designers work remotely leveraging their corresponding physical spaces but can collaborate in a single virtual environment where artifacts can be shared, compared, and reorganized together in real-time. Anchoring could also become temporal, with contents surfacing only when relevant and fading until they are needed again.

\subsection{Dimension 4: Facilitating Different Levels of Collaboration}

Participants envisioned MR tools as valuable for individual work, particularly during ideation and early exploration. They imagined MR as a personal creative space where sketches, references, and prototypes could be quickly externalized, rearranged, and tested without distraction. Such tools were seen as supporting both efficiency and creative freedom, enabling designers to duplicate design components, explore user flows, and manage early ideas with greater ease and flexibility than traditional tools, while maintaining control over their process before involving others. On the other hand, many participants imagined MR tools for collaborative work, where multiple designers and other stakeholders could contribute to the same design environment in real time or in remote collaboration settings. They saw opportunities for shared brainstorming, collective sketching, co-design workflows, and feedback collection sessions, where sketches could be annotated by others, ideas compared side by side, and contributions layered into evolving artifacts. 

The implications for future MR tools go beyond simply offering separate spaces for personal and group work. The deeper challenge lies in how these tools manage access to private physical space during collaboration~\cite{10.1145/3677386.3682085}. Unlike traditional digital tools (e.g., Figma), which only share what is explicitly placed on a canvas, MR tools often integrate digital content with the user's physical surroundings. This means collaborators may unintentionally see aspects of someone's personal space, such as sketches on a desk, objects in the room, or bodily gestures not intended to be shared. They may also manipulate or place digital content within another person's physical environment, introducing additional issues of trust and control. Addressing these challenges of spatial visibility and personal boundaries is essential for future MR tools. This tension is also reflected in recent MR systems such as the work done by \citet{10.1145/3613904.3642293}, which combines physical whiteboard interaction with reconfigurable shared virtual spaces, illustrating how collaborative environments must carefully balance physical grounding with flexible spatial reconfiguration.

\subsection{Limitations and Future Work}

While this study offers valuable insights into the promise of Mixed Reality (MR) as a bridge between physical and digital design environments, several limitations must be acknowledged. The number of participants was relatively small, which limits the generalizability of our findings across the broader community of UI/UX designers. Additionally, many of our participants were early-career designers, which may have influenced their perceptions of workflow flexibility and their openness toward adopting new MR-based design practices. While the experienced practitioners in our sample (P2 and P6) had similar responses, others may still perceive such tools differently. In addition, the MR probe we created was effective for sparking reflection but offered only a narrow view of possible interactions. This limitation may have constrained the range of ideas participants explored, although we deliberately kept the probe simple to minimize such effects. Moreover, the conceptual-probe study sessions were conducted individually and out of the participants' typical working environments, restricting opportunities to observe collaboration or work in natural environments. Further, most participants had little prior experience with MR, and their exposure to the conceptual probe was limited to a relatively short period of time. As a result, some of the perceived values toward MR may have been influenced by novelty and first-impression reactions.

These limitations point toward several promising directions for future research. Larger and more diverse studies would provide a fuller picture of MR's role in design practice. Equally important, group-based and remote co-design studies can help understand how MR might enhance collaboration across distributed teams. At the same time, future work should also prioritize accessibility and inclusiveness, ensuring that MR tools can be adapted to the needs of designers with varied abilities, resources, and working conditions.

\section{Conclusion}
In this paper, we explored how UI/UX designers envision Mixed Reality (MR) to integrate physical and digital design environments to create more seamless design workflows. Our participants valued MR for its ability to preserve the spontaneity and creative freedom of paper-based physical artifacts while extending the precision, organization, and collaborative capabilities of digital platforms. They also considered MR tools to have the potential to maintain workflow continuity across mediums, reduce manual effort in translating ideas between paper-based and digital artifacts, and support richer feedback and collaborative interactions. These findings contribute a set of four key design dimensions for future MR tools: (1) integrating diverse and heterogeneous artifacts, (2) blending lo-fi and hi-fi prototyping, (3) leveraging spatial anchoring, and (4) supporting different levels of collaboration. Together, these dimensions provide actionable guidance for researchers and practitioners, highlighting how MR can evolve into a powerful medium for design that combines the physical and digital environments to expand opportunities for creativity, efficiency, and collaboration.

\begin{acks}
We thank our participants for their time and valuable insights. This work is partially supported by the Canada Research Chairs program (CRC-2021-00076) and the Natural Sciences and Engineering Research Council of Canada (RGPIN-2018-04470).
\end{acks}

\balance
\bibliographystyle{ACM-Reference-Format}
\bibliography{references}

\end{document}